# HPTT: A High-Performance Tensor Transposition C++ Library

Paul Springer
AICES, RWTH Aachen University, Germany
springer@aices.rwth-aachen.de

Tong Su
RWTH Aachen University, Germany
tong.su@rwth-aachen.de

Paolo Bientinesi
AICES, RWTH Aachen University, Germany
pauldj@aices.rwth-aachen.de

## Abstract

Recently we presented TTC, a domain-specific compiler for tensor transpositions. Despite the fact that the performance of the generated code is nearly optimal, due to its offline nature, TTC cannot be utilized in all the application codes in which the tensor sizes and the necessary tensor permutations are determined at runtime. To overcome this limitation, we introduce the open-source C++ library **H**igh-**P**erformance **T**ensor **T**ransposition (HPTT). Similar to TTC, HPTT incorporates optimizations such as blocking, multi-threading, and explicit vectorization; furthermore it decomposes any transposition into multiple loops around a so called micro-kernel. This modular design—inspired by BLIS—makes HPTT easy to port to different architectures, by only replacing the hand-vectorized micro-kernel (e.g., a $4 \times 4$ transpose). HPTT also offers an optional autotuning framework—guided by performance heuristics—that explores a vast search space of implementations at runtime (similar to FFTW). Across a wide range of different tensor transpositions and architectures (e.g., Intel Ivy Bridge, ARMv7, IBM Power7), HPTT attains a bandwidth comparable to that of SAXPY, and yields remarkable speedups over Eigen's tensor transposition implementation. Most importantly, the integration of HPTT into the Cyclops Tensor Framework (CTF) improves the overall performance of tensor contractions by up to $3.1\times$.

*Categories and Subject Descriptors* D.1.3 [*Concurrent Programming*]: Parallel programming; G.4 [*Mathematical Software*]: Parallel and vector implementations; I.1 [*Symbolic and Algebraic Manipulation*]: Languages and Systems—Special-purpose algebraic systems

*Keywords* multidimensional transposition, High-Performance Computing, vectorization, tensors, autotuning

## 1. Introduction

Tensors, or multidimensional arrays, are ubiquitous in various scientific fields such as machine learning [1, 24], quantum chemistry calculations [2, 8], multidimensional Fourier transforms [5, 16] and climate simulations [4]. The manipulation of tensors, via operations such as transposition, contraction,[1] completion, and factor-ization, are performance critical tasks. This work introduces a high-performance library for tensor transpositions.

[1] A tensor contraction is the generalization of a matrix-matrix multiplication.

Transpositions are preparatory tasks that play a role within many other operations. For instance, tensor contractions can be cast in terms of general matrix-matrix multiplications (GEMM), an operation for which there exist numerous highly-optimized implementations (e.g., BLIS, OpenBLAS, MKL); however, the approach is only useful in combination with an efficient and flexible tensor transposition kernel. Despite the unfavourable memory access patterns that arise in high-dimensional transpositions, recent work [12, 20, 21, 26] demonstrated that a domain-specific compiler, such as our Tensor Transposition Compiler (TTC), is capable of generating highly-efficient routines, both for CPUs and accelerators, for a given transposition and problem sizes. While TTC delivers good performance, it is only applicable to transpositions for which the size and the required permutations are known at compile time. To overcome this issue, we designed **H**igh **P**erformance **T**ensor **T**ransposition (HPTT), an open-source C++ library for transpositions of the form

$$\mathcal{B}_{\Pi(i_1 i_2 \ldots i_N)} \leftarrow \alpha \times \mathcal{A}_{i_1 i_2 \ldots i_N} + \beta \times \mathcal{B}_{\Pi(i_1 i_2 \ldots i_N)}, \qquad (1)$$

where $\mathcal{A}$ and $\mathcal{B}$ are $N$-dimensional tensors, $\Pi(i_1 i_2 \ldots i_N)$ denotes an arbitrary permutation of the indices $i_1, i_2, \ldots, i_N$, and $\alpha$ and $\beta$ are scalars. This form enables HPTT to transpose $\mathcal{A}$ into $\mathcal{B}$, and also to scale either of the operands.

Throughout this publication we adopt the Fortran memory layout. Thus, storing the tensor indices from left to right; given an $N$-dimensional tensor $\mathcal{A}_{i_1 i_2 \ldots i_N}$, $i_1$ and $i_N$ respectively are the *fastest-varying* (also known as *stride-1*) and *slowest-varying* indices.

The remainder of this paper is structured as follows. Section 2 outlines related work. The design and structure of HPTT are discussed in Sec. 3, while Sec. 4 contains a performance evaluation, with a breakdown of the used optimization techniques. Conclusions are drawn in Sec. 5.

## 2. Related Work

Two-dimensional tensor transposition (i.e., matrix transposition) is a well studied operation, including optimizations for blocking, vectorization, unrolling, and software prefetching [3, 6, 11, 13, 14, 25]. The same optimizations are investigated in the context three-dimensional out-of-place tensor transpositions on CPUs [10, 22].

The optimization of arbitrary-dimensional tensor transpositions has gained more interest in recent years [12, 20, 21, 26].

Wei et al. [26] presented a code generator which "uses exhaustive global search", explores blocking, in-cache buffers to avoid conflict misses, loop unrolling, software prefetching and vectorization. While their implementation exhibits good performance on the selected architectures, neither parallelization, nor different loop orders have been considered.

Our previous work on the **T**ensor **T**ransposition **C**ompiler (TTC) [20, 21] relied on *code generation* to yield a nearly-optimal implementations for any given tensor transposition.

Lyakh et al. [12] designed a generic multidimensional transpose algorithm and evaluated it across different architectures (e.g., Intel Xeon, Intel Xeon Phi, AMD and NVIDIA K20X). Despite the fact that their algorithm outperforms a naive baseline implementation, the results suggest that there still exists a noticeable performance gap to the bandwidth attained by a direct copy.

The recently released CUDA Tensor Transpose (cuTT) library [9] provides high-performance tensor transpositions for NVIDIA GPUs. Despite the fact the HPTT and cuTT target different architectures, both libraries use similar concepts (e.g., an FFTW-like autotuning approach based on plans). Given the different platforms targeted by cuTT and HPTT, these libraries complement each other well.

## 3. High-Performance Tensor Transpositions

HPTT is an open-source[2] high-performance C++ library for out-of-place tensor transpositions running on shared-memory systems. It inherits its key design principle from TTC: any tensor transposition is decomposed into independent 2D tensor transpositions; each of these 2D transpositions is then computed by a so-called *macro-kernel* that is again broken down into loops around a (smaller) *micro-kernel* (see Section 3.2 for details). This design—inspired by BLIS [23]—allows HPTT to be easily ported to different architectures because only the micro-kernel needs to be manually implemented via vector intrinsics. Like TTC, HPTT still exhibits desirable properties such as (optional) autotuning (Section 3.1), multi-threading support (Section 3.2), and explicit vectorization (Section 3.2), yielding high performance across a wide range of tensor transpositions and architectures.

The key differences with respect to TTC are the following: First and foremost, HPTT does not require recompilation for different tensor transpositions and sizes; this makes HPTT applicable to scenarios where those parameters can only be determined at runtime (e.g., within CTF [18] or Eigen [7]). Second, HPTT is able to search for different *parallelization strategies* (see Section 3.2) and it avoids the search for different blocking sizes (see Section 3.2). Finally, in contrast to TTC, HPTT needs to perform the autotuning at runtime. To this end, we adopted a recursive design—much like BLIS [23]—which takes an additional parameter that encodes the execution process for any given tensor transposition, henceforth called *plan*.[3]

### 3.1 Plan-Creation and Autotuning

```
1  // corresponds to i_1
2  Plan *i1 = new Plan(0 /*start*/,  6 /*end*/, 1 /*inc*/,
3                     1 /*strideA*/, 4 /*strideB*/, NULL);
4  // corresponds to i_2
5  Plan *i2 = new Plan(0 /*start*/ , 4 /*end*/, 1 /*inc*/,
6                     6 /*strideA*/, 1 /*strideB*/, i1);
```

**Figure 1:** Plan data structure example for a single-threaded, unblocked 2D transposition: $\mathcal{B}_{i_2 i_1} \leftarrow \mathcal{A}_{i_1 i_2}, \mathcal{A} \in \mathbb{R}^{6\times 4}, \mathcal{B} \in \mathbb{R}^{4\times 6}$.

A plan encodes the execution of any given tensor transposition. As such, a plan represents all loops, each corresponding to a different tensor index. Figure 1 outlines the plan for a single-threaded, unblocked 2D transposition $\mathcal{B}_{i_2 i_1} \leftarrow \mathcal{A}_{i_1 i_2}, \mathcal{A} \in \mathbb{R}^{6\times 4}, \mathcal{B} \in \mathbb{R}^{4\times 6}$. The plans i1 and i2 respectively correspond to indices $i_1 \in \{0, 1, ..., 5\}$ and $i_2 \in \{0, 1, 2, 3\}$. More precisely, the strides of i1 with regard to $\mathcal{A}$ and $\mathcal{B}$ are 1 and 4, respectively. The increment (inc) is not important for now, since this plan corresponds to an unblocked implementation (we consider blocking in Section 3.2).

---

[2] Published under LGPLv3 at www.github.com/springer13/hptt.

[3] The plan is conceptually very similar to what is called a *control tree* (cntl_t) in BLIS.

The loop order is determined by the order i1 and i2 point to one another: i2 stores a pointer to i1 (Line 6), while i1's pointer is set to NULL (Line 3), indicating that no more indices follow and that the macro-kernel should be invoked. This example illustrates an execution where i1 and i2 respectively correspond to the fastest-varying and the slowest-varying index (i.e., the loop associated to $i_1$ is the innermost). A different loop order can be obtained by having i1 point to i2 and i2 point to NULL. This flexible design enables HPTT to generate different plans—based on different loop orders—quite easily at runtime; likewise, different *parallelization strategies* can be accommodated as well (see Section 3.2).

Figure 2 illustrates the *plan creation* process of HPTT. Given the input—in form of the permutation, size, number of threads and a timeout parameter (for autotuning)—HPTT begins by (1) merging indices which are consecutive in both tensors into a "superindex".[4] HPTT's plan creation process offers two different execution paths: The *autotune path* and the *quick path*. The former (2) generates and (3) ranks **all** plans followed by the autotuning process (4)-(6). During the *quick path* (denoted by the green dashed arrow), on the other hand, only a **single**, good plan is created and returned immediately without the need to generate all plans. The *quick path* is essential for keeping the overhead—due to the plan creation—as low as possible when no autotuning is desired (e.g., from within CTF).

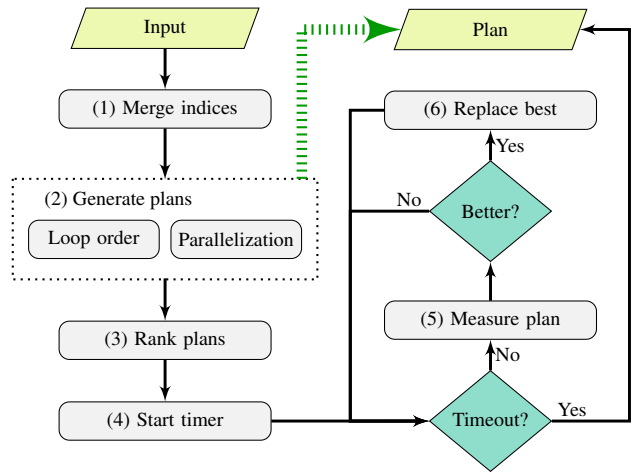

**Figure 2:** HPTT's plan creation process. The dashed green arrow denotes the *quick path* which returns a good plan immediately.

Whether HPTT takes the *autotune path* or the *quick path* depends on the user-specified timeout parameter; the user, thus, has to decide if the overhead due to the autotuning process can be amortized by repeated executions of the same tensor transposition (e.g., [17])—albeit with (possibly) different data. If the *au-*

---

[4] For instance, $\mathcal{B}_{i_2 i_3 i_1} \leftarrow \mathcal{A}_{i_1 i_2 i_3}$ becomes $\mathcal{B}_{(i_2 i_3) i_1} \leftarrow \mathcal{A}_{i_1 (i_2 i_3)}$, see [20].

*totune path* is chosen, then (4) a timer is started and HPTT evaluates the performance of the ranked plans (from good to bad) until the timeout has been reached. This autotuning feature is conceptually very similar to that of FFTW [5]. Like FFTW, HPTT also uses the original input data (i.e., the pointers to $\mathcal{A}$ and $\mathcal{B}$) for autotuning. However, in contrast to FFTW, HPTT does not modify the elements of either input during the autotuning process. We achieve this by setting the coefficient $\alpha = 0$ and $\beta = 1$; thus, the output tensor is only overwritten with its original value, while $\mathcal{A}$ is still read from main memory but it is then multiplied with zero.[5]

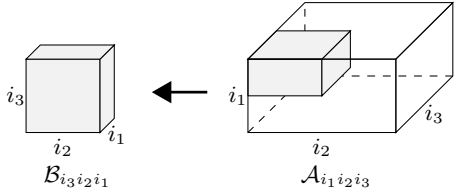

(a) Visualization.

```
1   // specify permutation and size
2   std::vector<uint32_t> perm    = {2,1,0};
3   std::vector<uint32_t> sizeA   =     { 8, 16, 16};
4   std::vector<uint32_t> outerSizeA = {16, 32, 32};
5   std::vector<uint32_t> outerSizeB = {16, 16,  8};
6
7   // create a plan
8   double timeout = 1.0; // in seconds
9   auto plan = hptt::create_plan(perm,
10                        1.0 /*alpha*/, A, sizeA, outerSizeA,
11                        0.0 /*beta*/ , B,        outerSizeB,
12                        numThreads, timeout);
13
14  // execute the transposition
15  plan->exec();
```

(b) HPTT input.

**Figure 3:** Exemplary tensor transposition $\mathcal{B}_{i_3 i_2 i_1} \leftarrow \mathcal{A}_{i_1 i_2 i_3}$ for $\mathcal{A} \in \mathbb{R}^{16 \times 32 \times 32}$, the shaded sub-tensor in $\mathcal{A}$ is of size $8 \times 16 \times 16$.

Figure 3 visualizes an exemplary use case of HPTT where a subtensor of $\mathcal{A} \in \mathbb{R}^{16 \times 32 \times 32}$ is transposed-and-compacted into $\mathcal{B} \in \mathbb{R}^{16 \times 16 \times 8}$ via the tensor transposition $\mathcal{B}_{i_3 i_2 i_1} \leftarrow \mathcal{A}_{i_1 i_2 i_3}$. HPTT accepts so called *outer sizes* (similar to the leading dimension in BLAS) which enable HPTT to operate on subtensors; this feature makes it possible to transpose-and-compact (see Figure 3a) or to transpose-and-scatter (reverse operation of Figure 3a, not shown).

### 3.2 Vectorization and Parallelization

HPTT decomposes an arbitrary-dimensional tensor transposition into many independent two-dimensional $b \times b$ *macro-tiles* which are composed of $w \times w$ *micro-tiles* (see Fig. 4). These two-dimensional tiles are, always chosen such that the stride-1 index in both $\mathcal{A}$ and $\mathcal{B}$ is preserved, facilitating fully-vectorized memory operations and thereby exploiting the spatial locality inherent to tensor transpositions. As illustrated in Fig. 4, the macro- and micro-tiles are respectively computed by a *macro-* and *micro-kernel*.

Each $w \times w$ micro-tile denote an explicitly vectorized, in-register transposition, with $w$ corresponding to the vector-width of the underlying architecture (e.g., $w = 8$ for single-precision elements on an AVX-enabled processor). The interested reader is referred to [20] for further details on the vectorization.

A noticeable difference to TTC [20, 21], however, is that the remainder (the macro-tiles shaded in gray, Fig. 4) are now also vectorized. HPTT starts out with $b = 4w$, once the remainder is reached $b$ is decreased to $b/2$; this concept continues until $b = w$. Thanks to this concept, HPTT does not search for suitable blocking parameters any longer and removes this search direction entirely, making the plan generation process less complex.

---

[5] $\mathcal{A}$ may not have special values like `inf` or `NaN`.

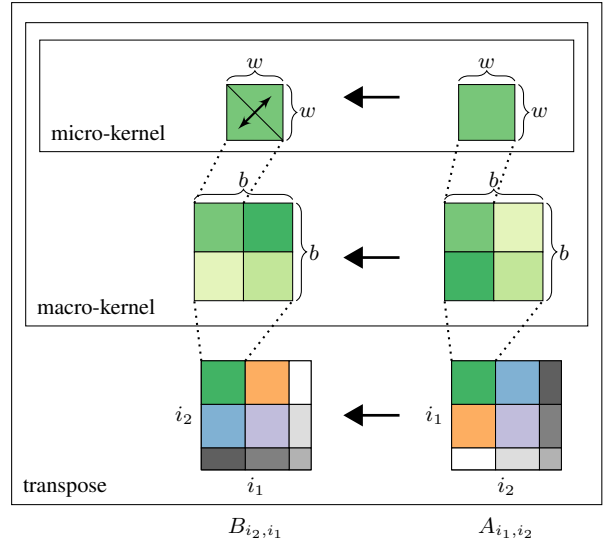

**Figure 4:** Decomposition of a 2D transposition into $b \times b$ macro-tiles and $w \times w$ micro-tiles.

The macro-tiles are completely independent from one another and can be computed in-parallel by different threads. HPTT is able to parallelize the loops, corresponding to the tensor indices, individually. For instance, revisiting Fig. 1, two threads $t_1$ and $t_2$ could parallelize the loop associated to $i_1$, if $t_1$ uses `i1->start=0; i1->end=3;` while $t_2$ uses `i1->start=3; i1->end=6;`; the loop corresponding to $i_2$ could also be parallelized similarly. Likewise, HPTT can parallelize multiple loops simultaneously; this mechanism enables a new search direction which had not been present in TTC before.

### 3.3 Performance Heuristics

Given the vast search space of viable plans, the challenge is to select a good plan that yields high performance. This sections outlines the performance heuristics of HPTT which addresses this problem. The performance model consist of two separate heuristics: the parallelization heuristic (see Section 3.3.1) and the loop order heuristic (see Section 3.3.2).

#### 3.3.1 Parallelization

As mentioned above, HPTT is able to parallelize all loops individually resulting in a large search space of different parallelization strategies.

The fundamental ideas behind the parallelization heuristic are manifold: First and foremost, load-balancing should be maximized; thus, total amount of work should be equally distributed among the threads. Second, the parallelization of any stride-1 index should be avoided; abiding to this rule increases the amount of consecutive memory accesses per thread. Finally, if one of the stride-1 indices needs to be parallelized in order to increase load-balancing, then this heuristic prefers to parallelize the stride-1 index of $\mathcal{A}$ over the stride-1 index of the output tensor $\mathcal{B}$; the rationale being that *false sharing* between the threads should be avoid. An analytical description of this heuristic is beyond the scope of this paper;

however, a curious reader can find the corresponding source code at www.github.com/springer13/hptt.

### 3.3.2 Loop Order

A tensor transposition of $N$-dimensional tensors has $N!$ distinct ways to order the loops. Most of these loop orders lead to significantly different performance [20], making it critical to choose a good loop order for any given tensor transposition.

We encode a loop order $L$ of an $N$-dimensional tensor transposition as an $N$-tuple $L = (l_1, l_2, ..., l_N)$ with $l_i \in \{1, 2, ..., N\}$ such that the loop corresponding to index $l_i$ represents the $i$-th loop around the macro-kernel. For instance, given the tensor transposition $\mathcal{B}_{i_6 i_5 i_4 i_1 i_3 i_2} \leftarrow \mathcal{A}_{i_1 i_2 i_3 i_4 i_5 i_6}$ and the loop order $L = (6, 5, 4, 1, 3, 2)$ means that any *plan* that uses $L$ traverses $\mathcal{B}$ in a linear fashion. While such a loop order is ideal for $\mathcal{B}$, it would (most likely) be supoptimal with respect to the memory accesses to $\mathcal{A}$. More precisely, the loop corresponding to the stride-1 index of $\mathcal{A}$ ($i_1$) is the fourth loop around the macro-kernel, leading to a strided memory access pattern with a large stride.

The rationale behind HPTT's loop heuristic is that loops corresponding to innermost indices of either of $\mathcal{A}$ and $\mathcal{B}$ should be "close" to the macro-kernel. Moreover, this heuristic slightly favours the innermost indices of $\mathcal{B}$ over those of $\mathcal{A}$ to favour consecutive writes over consecutive reads.

|      | Permutation |         |         |
|------|-------------|---------|---------|
| Rank | 6,5,4,3,2,1 | 4,3,6,2,1,5 | 6,1,5,4,3,2 |
| 1 | 6,1,5,2,4,3 | 4,1,3,2,6,5 | 1,6,2,5,3,4 |
| 2 | 6,1,2,5,4,3 | 1,4,3,2,6,5 | 1,6,2,3,5,4 |
| 3 | 1,6,5,2,4,3 | 4,1,2,3,6,5 | 1,6,5,2,3,4 |
|   | (a) | (a) | (c) |

**Table 1:** Top-3 loop orders ranked by the loop heuristic for three different permutations. The leftmost index denotes the innermost loop, while the rightmost index corresponds to the outermost loop.

Table 1 shows the top-3 loop orders for three different tensor transpositions (a) – (c). We observe that the stride-1 indices are always either the fastest-varying (innermost) or the second fastest-varying index. For instance, the top-1 loop order for the tensor transposition $\mathcal{B}_{i_6 i_5 i_4 i_3 i_2 i_1} \leftarrow \mathcal{A}_{i_1 i_2 i_3 i_4 i_5 i_6}$ in column (a) interleaves the indices of $\mathcal{B}$ and $\mathcal{A}$ while given precedence to $\mathcal{B}$. On other hand, the top-1 loop order for the tensor transposition $\mathcal{B}_{i_6 i_1 i_5 i_4 i_3 i_2} \leftarrow \mathcal{A}_{i_1 i_2 i_3 i_4 i_5 i_6}$ in column (c) chooses $i_1$ as the innermost index due to the fact that $i_1$ is the second fastest-varying index in $\mathcal{B}$ while $i_6$ (the fastest-varying index of $\mathcal{B}$) is the slowest-varying index (i.e., least important) index of $\mathcal{A}$; phrased differently, $i_1$ is important to both $\mathcal{A}$ and $\mathcal{B}$, while $i_6$ is only important to $\mathcal{B}$.

### 3.4 Recursive Design

Figure 5 illustrates the recursive design of HPTT. Each invocation of `transpose()` either invokes the macro-kernel (Lines 7-9) or enters a loop and then unfolds the plan (Lines 12-14) with a recursive call. In the latter case, the pointers to $\mathcal{A}$ and $\mathcal{B}$ are offset according to the strides of the respective tensor index and the plan is unfolded (Line 14) before making the recursive.

## 4. Performance Evaluation

To assess the performance of HPTT across a wide range of use-cases, we report the bandwidth attained on a tensor transpositions benchmark [20] that comprises a total of 57 transpositions ranging from 2D to 6D, with each tensor roughly occupying 200 MB of memory. All measurements are based on the maximum bandwidth attained over multiple executions (caches are cleared in between runs). If not otherwise noted, we use single-precision tensors, initialized in a NUMA-friendly fashion to distribute them

```
void transpose( const float* A, float alpha,
                float* B, float beta, const Plan* plan){
    int strideA = plan->strideA;
    int strideB = plan->strideB;
    if( isMacroKernel(plan) ){
        /*** invoke macro-kernel ***/
        for(int i = plan->start; i < end; i+= plan->inc)
            macroKernel(&A[i*strideA], alpha, plan->next->strideA,
                        &B[i*strideB], beta,  plan->next->strideB);
    } else {
        /*** recurse ***/
        for(int i = plan->start; i < plan->end; i+= plan->inc)
            transpose( &A[i*strideA], alpha,
                       &B[i*strideB], beta, plan->next);
    }
}
```

**Figure 5:** Plan Execution. This function has been simplified from its original form for better readability.

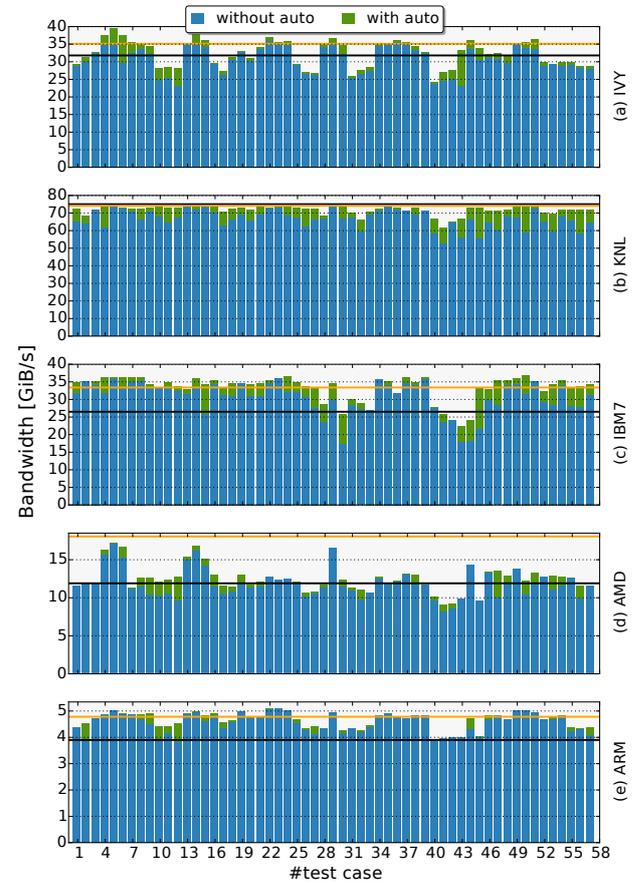

**Figure 6:** HPTT bandwidth with and without autotuning. Horizontal orange and black lines respectively denote the system's SAXPY and STREAM bandwidth.

evenly among the memory controllers. The bandwidth is calculated as

$$\text{Bandwidth} := \frac{\lambda \times S}{1024^3 \times \text{Time}} \; [\text{GiB/s}], \quad (2)$$

where $S$ denotes the size of the transposed tensor (in bytes) and $\lambda$ is either 2 or 3, depending on whether $B$ is overwritten ($\beta = 0$) or updated ($\beta \neq 0$); unless otherwise stated, we report the results for $\beta \neq 0$.

| Name | Microarchitecture | Model | #Cores | #Threads | Bandwidth [GiB/s] | | Compiler | Compiler-flags |
|---|---|---|---|---|---|---|---|---|
| | | | | | SAXPY | STREAM | | |
| IVY | Ivy Bridge | Intel E5-2670 v2 | $2 \times 10$ | $2 \times 10$ | 35.1 | 31.8 | icpc 16.0.3 | -O3 -xHost |
| KNL | Knights Landing | Intel Xeon Phi 7210 | 64 | 64 | 74.1 | 75.0 | icpc 17.0.2 | -O3 -xHost |
| IBM | Power7 | IBM PowerPC A2 | 16 | 64 | 33.4 | 26.5 | g++ 6.3 | -O3 -mcpu=native |
| STR | AMD Steamroller | A10-7850K | 4 | 4 | 18.1 | 11.9 | g++ 5.3 | -O3 -march=native |
| ARM | ARMv7-A | ODROID-XU3 | 4 + 4 | 8 | 4.8 | 3.9 | g++ 5.4 | -O3 -march=native |

**Table 2:** Hardware description. The *SAXPY* and STREAM columns indicate the bandwidth attained for $\mathbf{y} \leftarrow \alpha\mathbf{x}+\mathbf{y}$, and for the $\mathbf{z} \leftarrow \alpha\mathbf{x}+\mathbf{y}$ (STREAM *triad* [15]), respectively.

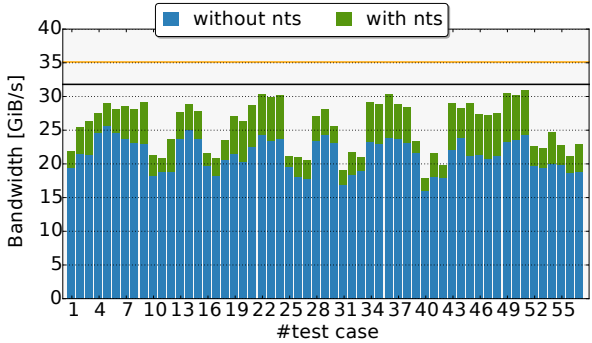

**Figure 7:** HPTT bandwidth for $\beta = 0$ on Ivy Bridge system. Horizontal orange and black lines denote the system's SAXPY and STREAM bandwidth, respectively.

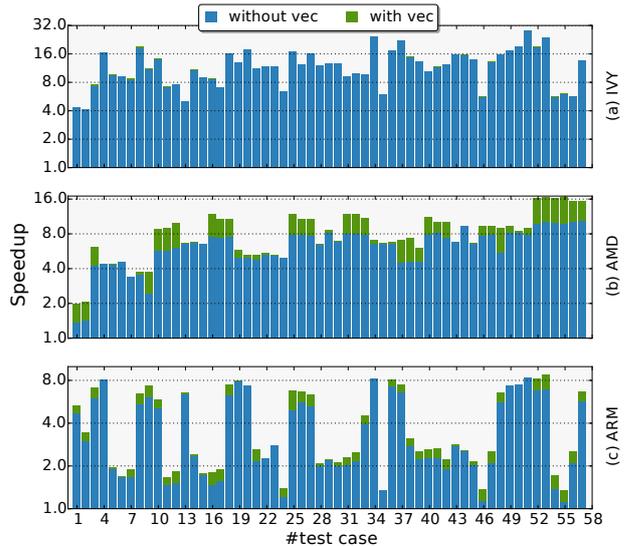

**Figure 8:** HPTT speedup over Eigen with and without explicit vectorization using a semi-log scale.

We evaluate HPTT's performance on five platforms (Table 2); the number of threads, per platform, were chosen empirically to give the best performance.

In Fig. 6, we observe that (1) on average, HPTT attains the following fraction of the SAXPY / STREAM bandwidth for the various systems: 92% / 102% (IVY), 97% / 96% (KNL), 100% / 126% (IBM), 69% / 105% (AMD) and 97% / 119% (ARM); these results illustrate HPTT's close-to-optimal performance. (2) Autotuning, on average, only improves the performance by 6%, 8%, 10%, 6%, and 2% for the IVY, KNL, IBM, AMD and ARM system, respectively; these numbers indicate that the performance heuristics work well in most cases. (3) The maximum speedup due to autotuning can be as high as $1.45\times$, $1.31\times$, $1.53\times$, $1.35\times$ and $1.19\times$ for the IVY, KNL, IBM, AMD and ARM system, respectively; thus, making autotuning a viable option for those cases where the autotuning overhead can be amortized over several tensor transpositions.

Figure 7 covers the case $\beta = 0$ (i.e., an out-of-place transposition without accumulating into the output) on the IVY system, and highlights the impact of non-temporal stores. While the overall performance is lower than what outlined in Fig. 6 ($\beta \neq 0$), on average non-temporal stores improve the performance by $1.20\times$.

HPTT's speedup over Eigen [7], with and without explicit vectorization, are shown in Fig. 8. The maximum speedup for the IVY, AMD, and ARM systems is as high as $27.4\times$, $16.6\times$, and $8.8\times$, respectively. Explicit vectorization does not affect the performance on the Ivy Bridge (or and KNL systems; data not shown); we take this as a clear indication of icpc's superb optimization capabilities. Vice versa, both the AMD and ARM systems experience noticeable average speedups of $1.28\times$ and $1.12\times$ due to explicit vectorization, suggesting that gcc struggles to find a good vectorization.

### 4.1 Cyclops Tensor Framework

We analyze now the effects of HPTT on CTF [18].[6] Since one tensor contraction within CTF can require up to six tensor transpositions, highly-optimized tensor transpositions are critical to achieve good overall performance. We measure the speedup over CTF for a tensor contraction $\chi$ as

$$\text{Speedup}(\chi) = \frac{\text{Time( CTF without HPTT}(\chi))}{\text{Time( CTF with HPTT}(\chi))}. \quad (3)$$

Figure 9 shows that HPTT improves CTF's performance noticeably; the test cases are sorted identically to those presented in [19], that is, bandwidth-bound and compute-bound contractions on the left and right, respectively. As expected, we observe larger speedups (up to $3.1\times$) for those contractions limited by memory bandwidth, than for the compute-bound ones. This is intuitive because the contractions towards the left spend more time on transpositions than those towards the right, which are instead dominated by a matrix-matrix multiplication (GEMM); test cases 13 and 19 are indeed pure GEMMs and thus do not require any transposition. Finally, we stress that these timings also include the time to create the "best" plan (according to HPTT's performance heuristics); the plan creation overhead is—thanks to the *quick path* (see Section 3.1)—negligible (on average, less than 0.1%).

---

[6] While CTF targets distributed-memory systems, it uses a shared-memory tensor transposition (via OpenMP) per MPI rank.

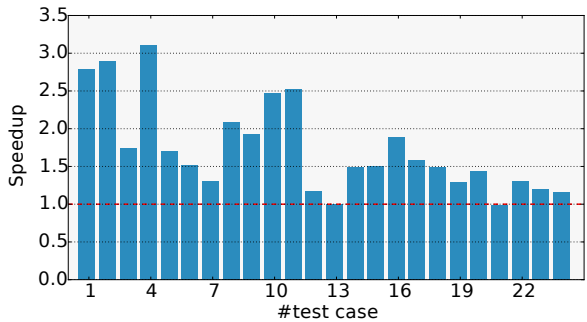

**Figure 9:** HPTT's impact on CTF's performance using double precision on the IVY system.

## 5. Conclusions and Future Directions

We introduced HPTT, an open-source[7] C++ library for high-performance tensor transpositions that avoids the code generation process of its predecessor, TTC [20]. At the same time, HPTT still preserves desirable properties such as autotuning, explicit vectorization, blocking, and multi-threading.

HPTT's autotuning framework is able to generate multiple *plans*; this feature enables HPTT to tune for an optimal plan with respect to any given tensor transposition and size at runtime. We further outlined HPTT's performance heuristics which yield competitive performance to the best plan which was found by the autotuning framework.

HPTT's close-to-optimal performance was demonstrated on a wide range of architectures, suggesting that there is essentially no more performance to be gained. We also assessed the performance of Eigen's tensor transposition implementation, indicated that HPTT executes up to $27.4\times$ faster. Given Eigen's importance for TensorFlow [1], HPTT could also proof to be valuable within the machine learning community.

Provided HPTT's close-to-optimal performance and its rich feature set (e.g., scaling, support for subtensors), this publication concludes our work on tensor transpositions. In the future it would be interesting to see if HPTT's autotuning process can also be applied to the BLIS framework.

Finally, we integrated HPTT into our local copy of CTF, resulting in an average speedup of $1.9\times$ across a wide range of tensor contractions.

### Acknowledgments

Financial support from the Deutsche Forschungsgemeinschaft (DFG) through grant GSC 111 is gratefully acknowledged. Furthermore, we thank Edgar Solomonik for assisting us to integrate HPTT into CTF. We also greatly appreciate the valuable discussions with Field G. Van Zee regarding BLIS' internal design. We also thank Elmar Peise for his helpful feedback to this paper.

---

[7] Available at www.github.com/springer13/hptt.